# Synthesis, Structural and Optical Properties of (ALa)(FeMn)O$_6$ (A = Ba and Sr) Double Perovskites

Dinesh Kumar[a)], V. Sudarshan[b)] and Akhilesh Kumar Singh[c)*]

*School of Materials Science and Technology, Indian Institute of Technology (BHU), Varanasi, Uttar Pradesh-221005, India*

Email: [a)]dineshiitbhu@gmail.com, [b)]vaddugandla.sudarshan.mst12@itbhu.ac.in, [c)*]aksingh.mst@iitbhu.ac.in

**Abstract.** Here, we report structural and optical properties of (ALa)(FeMn)O$_6$ (A = Ba and Sr) double perovskite synthesized via auto-combustion followed by a calcination process. Rietveld refinement of the structure using x-ray diffraction data reveals that (BaLa)(FeMn)O$_6$ crystallizes into cubic crystal structure with space group Pm-3m while (SrLa)(FeMn)O$_6$ crystallizes into rhombohedral crystal structure having space group R-3c. The absorption spectrum measurement using UV-Vis spectroscopy reveals that these samples are perfect insulator having an energy band gap between conduction and valence band of the order of 6 eV.

**Keywords:** Perovskite; Rietveld refinement; X-ray diffraction; Auto-combustion; Band gap.

## INTRODUCTION

The perovskite structure having formula ABO$_3$ with cubic structure covers enormous variety of compounds; which can accommodate variety of cations at A-site and B-site simultaneously [1]. Adding one more cations at A-site and/or B-site may convert it into double perovskites AA'BB'O$_6$ or A$_2$BB'O$_6$ type which are more popular in recent years. The perovskite structured compounds can exhibit variety of properties such as multiferroic nature [2], piezoelectricity, ferroelectricity and non-linear optical behavior [3], high-temperature superconductivity [4-6], Griffiths' Phase like behavior, exchange bias effect [7,10] etc. The room temperature discovery of colossal magnetoresistance (CMR) in Sr$_2$FeMoO$_6$ double perovskite [8] has led to a renovated interest in double perovskite materials. Shaheen et al. [9] reported synthesis and structural analysis of ALaMnFeO$_6$ (A = Ba, Ca, Sr) double perovskites using solid state method. From their neutron diffraction analysis, they reported orthorhombic perovskite structure (space group Pbnm) for BaLaMnFeO$_6$ & CaLaMnFeO$_6$ and rhombohedral crystal structure (space groupR-3c) for SrLaFeMnO$_6$. In contrast, Ramesha et al. [13] suggested that BaLaFeMnO$_6$ and SrLaFeMnO$_6$ crystallize into cubic crystal structure (space group Pm-3m). More recently, Palakkal et al. [10] synthesized La$_2$CrMnO$_6$ and using X-ray powder diffraction studies reported that La$_2$CrMnO$_6$ crystallize into orthorhombic symmetry with space group Pbnm and shows multiple phase transition.

In this paper, we report the structural analysis and optical properties of BaLaFeMnO$_6$ (BLFMO) and SrLaFeMnO$_6$ (SLFMO) using X-ray diffraction and UV-Vis spectroscopy, respectively. Samples were synthesized by auto-combustion method.

## EXPERIMENTAL DETAILS

The polycrystalline ALaFeMnO$_6$ (A = Ba and Sr) double perovskites have been synthesized by auto-combustion method [11]. The sample BaLaFeMnO$_6$ is denoted by BLFMO while sample SrLaFeMnO$_6$ is denoted by SLFMO in this article. The stoichiometric amount of La$_2$O$_3$ (99.9%, Sigma Aldrich), SrCO$_3$ (98.0%, Sigma Aldrich), BaCO$_3$ (99.0%, Sigma Aldrich), Mn(CH$_3$COO)$_2$.4H$_2$O (99.0%, Sigma Aldrich), Fe(NO$_3$)$_2$.9H$_2$O (98.0%, Sigma Aldrich) were used for synthesis. Glycine (99.5%, Hi-media) was used as fuel during synthesis. La$_2$O$_3$, SrCO$_3$ and BaCO$_3$ were dissolved in dil. HNO$_3$, while Mn(CH$_3$COO)$_2$.4H$_2$O, Fe(NO$_3$)$_2$.9H$_2$O and glycine were dissolved in distilled water to prepare the precursor solutions. All the precursor solutions were mixed together in a large beaker and kept on magnetic hot plate under continuous stirring at 250°C. After 6-8h of continuous stirring solution gets thicker and auto-ignition occurs resulting huge amount of gasses and turned into blackish-brown powder. The obtained powder was calcined at 1300°C for 6h to form single pure phase of the compounds. The X-ray diffraction (XRD) measurement was performed on both samples using X-ray diffractometer (Rigaku, Miniflex 600) in the θ-2θ configuration with a CuK$_\alpha$ source over the 2θ range of 20°-120° at the 2θ steps of 0.02°. The structural analysis of the synthesized samples was performed by Rietveld refinement using FullProf Suite software package [12]. The energy band gap of the samples was measured using UV-Vis spectroscopy.

## RESULTS AND DISCUSSION

### Crystal Structure: Rietveld Refinement

The structure refinement for both the samples was carried out considering various possible structures. The room temperature powder XRD plots for BLFMO and SLFMO are depicted in Figure 1. The XRD pattern of BLFMO can be indexed with cubic crystal structure space group Pm-3m [9] while the XRD pattern of SLFMO can be indexed using rhombohedral crystal structure with space group R-3c [13]. Following earlier convention, hexagonal unit cell was used to refine the rhombohedral structure of $SrLaFeMnO_6$. In the process of Rietveld refinement, we consider occupancy of $Ba^{2+}/La^{3+}$ ions at 1(a) (0, 0, 0), $Fe^{3+}/Mn^{4+}$ ions at 1(b) (1/2, 1/2, 1/2) and $O^{2-}$ ions at 3(c) (1/2, 1/2, 0) for BLFMO, while for SLFMO, $Sr^{2+}/La^{3+}$ ions are occupied at 6a (0, 0, 3/4), $Fe^{3+}/Mn^{4+}$ ions at 6b (0, 0, 0) and $O^{2-}$ ions at 18e (x, y=x, 1/4). The insets of the Figure 1 demonstrate some selected peaks for BLFMO and SLFMO in 2θ range 76°-79° and 94°-98° clearly showing difference in cubic and rhombohedral structures of the two compounds. All the XRD peaks are singlet for BLFMO while non-cubic splittings corresponding to R-3c is seen for SLFMO. Please note that both the $K_{\alpha1}$ and $K_{\alpha2}$ reflections are present in the XRD patterns making doublets.

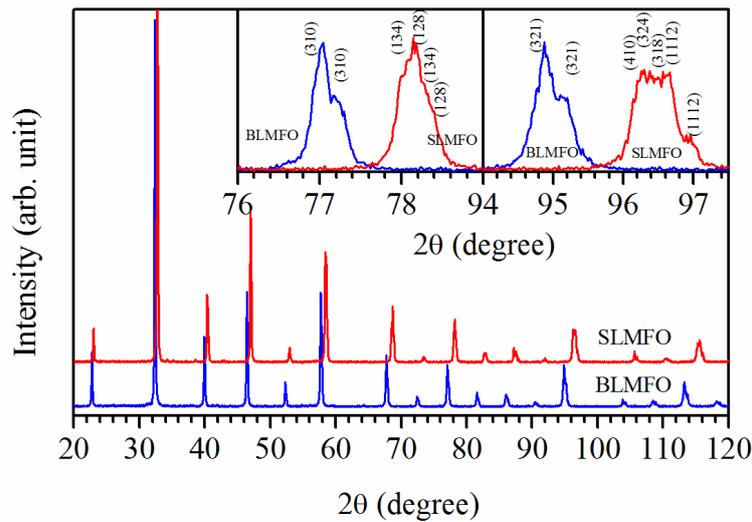

**FIGURE 1:** Room temperature XRD patterns of BLFMO and SLFMO. Insets show selected Bragg's reflections in 2θ between 76°-79° and 94°-98° for BLFMO and SLFMO.

Figures 2(a) & (b) show Rietveld fits for both the samples. The Rietveld fits are very good between observed and calculated patterns. The agreement factors for Rietveld fit are $R_B$ = 4.90 %, $R_f$ = 4.63 % & $\chi^2$ = 1.74 for BLFMO and $R_B$ = 3.88 %, $R_f$ = 4.61 % & $\chi^2$ = 1.54 for SLFMO. The dots in the figure represent observed XRD pattern while calculated pattern is shown by continuous curve overlapping to observed pattern. The difference profile between observed and calculated patterns is shown by continuous line. Whereas, the vertical bars just above difference profile indicate position of Bragg's reflections in the Rietveld fits. The lattice constant "a" and unit cell volume "V" for BLFMO (a = 3.91498(4) Å and V = 60.005(1) Å$^3$) is found to be consistent with the lattice constant reported by earlier authors (a = 3.916(1) Å) [13]. The hexagonal unit cell lattice parameters "a" and "c" and unit cell volume "V" are found to be 5.4759(1) Å, 13.3596(4) Å and 346.925(15) Å$^3$ for SLFMO, respectively. The refined crystallographic structural parameters, average bond-lengths and structural tolerance factor (t) for BLFMO and SLFMO are listed in Table1.

TABLE 1. Structural parameters, average bond-lengths and structural tolerance factor for BLFMO and SLFMO obtained after Rietveld analysis of the XRD patterns.

| Sample | a (Å) | c (Å) | V (Å³) | $\chi^2$ | $<d_{La-O}>$ | $<d_{Mn-O}>$ | t |
|---|---|---|---|---|---|---|---|
| BLFMO | 3.91498(4) | 3.91498(4) | 60.005(1) | 1.74 | 2.76830(2) | 1.95749(2) | 0.9999 |
| SLFMO | 5.4759(1) | 13.3596(4) | 346.925(15) | 1.54 | 2.737(5) | 1.940(6) | 0.9978 |

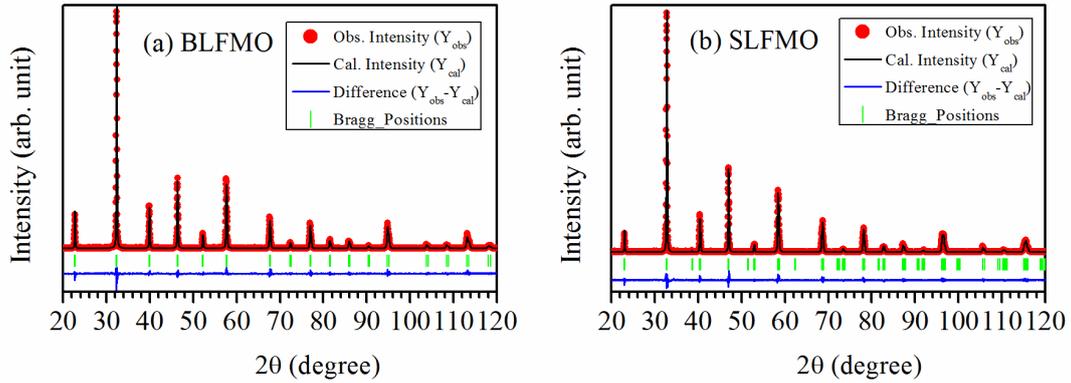

**FIGURE 2:** Rietveld fits for (a) BLFMO and (b) SLFMO.

## Optical Band-gap: UV-Vis absorption Spectroscopy

The UV-vis absorption spectra of the samples BLFMO and SLFMO recorded in the wavelength range of 200-800 nm and are shown in Figure 3(a). We have calculated the optical band gap for both the samples BLFMO and SLFMO by Wood and Tauc formula [14]:

$$\alpha h\nu = A(h\nu - E_g)^n \quad \ldots\ldots(1)$$

where, $h\nu$ is the energy of incident photon, $E_g$ is the optical band gap energy, A is a band tailoring constant and n is exponent. The value of exponent 'n' is 1/2 for direct band gap allowed transition. For calculation of direct optical band gap, the $(\alpha h\nu)^2$ versus $h\nu$ curves were plotted for BLFMO and SLFMO double perovskites and they are depicted in Figure 3(b). The direct optical band gap has been calculated and is found to be 5.88 eV for BLFMO and 5.86 eV for SLFMO double perovskites. The observed values of optical band gap are higher than 5 eV, which implies that the BLFMO and SLFMO double perovskites are highly insulator. Thus, these samples can be used in electronics where wide range band gap materials are required.

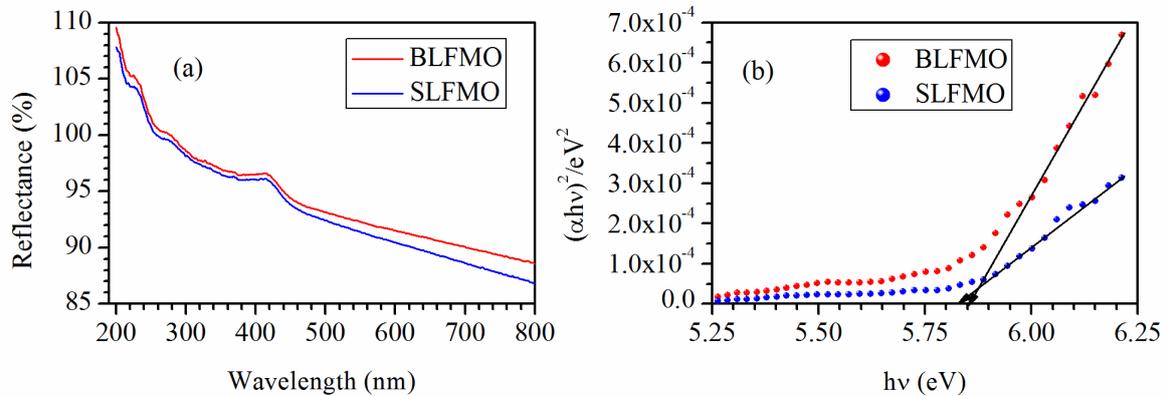

**FIGURE 3:** (a) UV-Vis absorption spectra of BLFMO & SLFMO and (b) Optical band gap measurements using $(\alpha h\nu)^2$ versus $h\nu$ curves for BLFMO & SLFMO.

## CONCLUSIONS

The double perovskites (ALa)(FeMn)$O_6$ (A = Ba and Sr) have been synthesized using auto-combustion synthesis method. The Rietveld refinement of structure using XRD patterns of the samples confirms cubic (Pm-3m) structure for BLFMO and rhombohedral R-3c) structure for SLFMO. The calculation of optical band gap for both the samples reveals highly insulating nature. The value of optical band gap has been found to be 5.88 eV and 5.86 eV for BLFMO and SLFMO, respectively.